\documentclass[osajnl,showpacs]{revtex4}  
 \usepackage{epsfig}

 \def\BEA {\begin{eqnarray}}
 \def\EEA {\end{eqnarray}}
 \def\BE {\begin{equation}}
 \def\EE {\end{equation}}
 \def\BA {\begin{array}}
 \def\EA {\end{array}}
 \def\NN {\nonumber}
 \def\re {\rm{Re} \,}
 \def\im {\rm{Im} \,}
  
\begin{document}

 \title{Tripartite entanglement in parametric down-conversion with
spatially-structured pump}

 \author{D.~Daems$^{1}$, F.~Bernard$^1$,  N.~J.~Cerf$^1$ and  M.~I.~Kolobov$^{1,2}$}
 \address{$^1$QuIC, Ecole Polytechnique, Universit\'e Libre de Bruxelles,
1050 Brussels, Belgium}
 \address{$^2$Laboratoire PhLAM, Universit\'e de Lille 1,
 59655 Villeneuve d'Ascq, France}

 \begin{abstract}
Most investigations of  multipartite entanglement have been
concerned with temporal modes of the electromagnetic field, and have
neglected its spatial structure.
We present a simple model which allows to generate
tripartite entanglement  between spatial modes by parametric
down-conversion with two symmetrically-tilted plane waves serving as a
pump. The
characteristics of this entanglement are investigated. We  also discuss the generalization of our scheme to $2N+1$-partite entanglement using
$2N$ symmetrically-tilted plane pump waves. Another interesting feature
 is the possibility of {\it entanglement localization} in just 
two spatial modes.
 \end{abstract}
 \pacs{270.5585   270.6570}
 \maketitle
 
 \section{Introduction}
Multipartite entanglement, as the name suggests, is the  entanglement
between more than two parties. The existence of this type of entanglement for
both discrete and continuous quantum variables is one of the most striking
predictions of quantum mechanics and is very unusual from the point of
view of classical physics. It is therefore not surprising that nowadays
multipartite entanglement is a very active research area of quantum optics
and quantum information. The interest to multipartite entanglement is
motivated not only by its fundamental nature but also by its potential for
applications in quantum communication technologies.

Numerous theoretical publications have addressed the 
characterization of multipartite entanglement.
Coffman, Kundu and Wootters [1] have established for a three-qubit system and
conjectured for $N$-qubit systems the so-called monogamy of quantum
entanglement, constraining the maximum entanglement between partitions of
a multiparty system. Later on, Adesso and Illuminati [2]-[5] have introduced the
continuous-variable tangle as a measure of multipartite entanglement for
continuous variable (CV) multimode Gaussian states and have demonstrated
that it satisfies the Coffman-Kundu-Wootters monogamy inequality.
The conjecture of Ref.~[1] has been proven by Osborne and Verstraete [6]. The corresponding proof for Gaussian states is in Hiroshima, Adesso and Illuminati [7].

Among the potential applications of CV multipartite entanglement let us
mention quantum teleportation networks, quantum telecloning, controlled
quantum dense coding, and quantum secret sharing [8].

Several generation schemes for CV multipartite entanglement have
been proposed in the literature theoretically and realized experimentally
in recent years. Among the first ones was the ``passive'' optical
scheme using squeezed states mixed with beam
splitters [9]. Then came the ``active'' schemes in which
multipartite entanglement is created as a result of parametric interaction
of several optical waves such as
cascaded/concurrent [10]-[12],
interlinked [13], [14] or consecutive parametric
interactions [15].
All of these schemes for generation of  CV multipartite entanglement are
considering the temporal modes of the electromagnetic field and neglect
its spatial structure. The question arises on whether the spatial modes
can also serve for the creation of CV multipartite entanglement [16]-[19]. 
In Ref.~[3], a physical implementation is given in terms of 2$N$-1 beam splitters and 2 $N$ single-mode squeezed input states, itself based on a previous work [20].

In this paper
we propose a simple scheme which corresponds to the ``active''
creation of tripartite entanglement between spatial modes of the
electromagnetic field in the process of parametric interaction.
We create spatial tripartite entanglement by pumping a
nonlinear parametric medium by a coherent combination of several tilted
plane monochromatic waves, what we call a spatially-structured pump. Since
the pump photon can be extracted from any of these waves and the pair of
down-converted photons emitted in different directions according to the
phase-matching condition, our scheme allows for the creation of tripartite
entanglement between spatial modes of the down-converted field. An interesting feature of our scheme is the
possibility of localizing the created spatial tripartite entanglement
in just two well-defined spatial modes formed as a linear combination of all
the modes participating in the down-conversion process.

In Sec. II, we shall consider explicitly the case of tripartite entanglement which is produced by parametric down-conversion with two symmetrically-tilted
plane waves serving as a pump. The genuine tripartite entanglement is analyzed in subsection A and its localization on two modes is studied in subsection B. We then discuss in Sec. III a generalization of our scheme to
$2N+1$-partite entanglement using $2N$ symmetrically-tilted pump waves and draw some conclusions.

\section{Spatial tripartite entanglement} 
\subsection{Parametric down-conversion with two symmetrically-tilted
plane waves} 

The scheme of parametric down-conversion used here is shown in Fig.~1. 
\begin{center}
\begin{figure}
\includegraphics[scale=0.54]{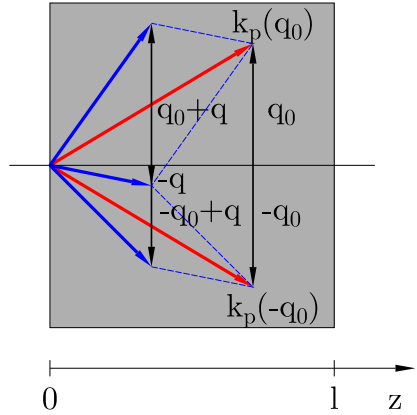}
\caption{(Color online) Scheme for generating tripartite entanglement between
spatial modes. The tilted pumps have wave vectors ${\bf k}_p (\pm q_0)$. The transverse (vertical) components are $\pm q_0$.}
\end{figure}
\end{center}
We consider two plane
pump waves with wave vectors ${\bf k}_p (\pm q_0)$  where $\pm q_0$ are the pertaining transverse components (which are vertical on Fig. 1)
at the input to the crystal. The pump field reads then
\BE
    E_p(x)=\alpha\Bigl(e^{iq_0 x}+e^{-iq_0 x}\Bigr) ,
    \EE
with $\alpha$  the amplitude of the pump, chosen as a real number in our case.
We shall denote the corresponding spatial Fourier components of
the corresponding field operators as follows
\BEA
 \hat{a}(z,q) &\equiv& \hat{a}_0(z)\NN\\
 \hat{a}(z,\pm q_0+q) &\equiv& \hat{a}_\pm(z). \label{mode0pm}
\EEA
These modes are depicted in  Fig.~1: the long (blue) arrows pertains to $\hat{a}_\pm(0)$, the short one to $\hat{a}_0^\dag(0)$.

 Due to the parametric interaction in the nonlinear medium these operators are
coupled and obey the following equations in the rotating wave approximation [21]:
\begin{eqnarray}
     \frac{d}{dz}\hat{a}_0 & = & \alpha \lambda \Bigl(\hat{a}_+^\dag
     +\hat{a}_-^\dag\Bigr)   e^{i\Delta z},
                 \nonumber\\
     \frac{d}{dz}\hat{a}_+ & = & \alpha \lambda \hat{a}_0^\dag e^{i\Delta z},
                 \nonumber\\
     \frac{d}{dz}\hat{a}_- & = & \alpha \lambda \hat{a}_0^\dag e^{i\Delta z},
                  \label{3eq}
\end{eqnarray}
where $\lambda$ is the coupling constant and $\Delta$ is the phase
mismatch.

The solution to this set of equations for the fields at the output of the
crystal, $z=l$, can be written analytically as follows
 \begin{eqnarray}
  \hat{a}_+(l) &=&
  \frac{1}{2}[U(l)+1]\hat{a}_+(0)+\frac{1}{2}[U(l)-1]\hat{a}_-(0)+
  \frac{1}{\sqrt{2}}V(l)\hat{a}_0^{\dag}(0),
        \nonumber\\
  \hat{a}_-(l) &=&
  \frac{1}{2}[U(l)-1]\hat{a}_+(0)+\frac{1}{2}[U(l)+1]\hat{a}_-(0)+
  \frac{1}{\sqrt{2}}V(l)\hat{a}_0^{\dag}(0),
        \nonumber\\
  \hat{a}_0(l) &=&
  U(l)\hat{a}_0(0)+\frac{1}{\sqrt{2}}V(l)[\hat{a}_+^{\dag}(0)+
  \hat{a}_-^{\dag}(0)],
        \label{solution}
 \end{eqnarray}
with the functions 
\BEA
U(z)&=&e^{i \Delta z/2} \left( \cosh(\Gamma z)
     - i\frac{\Delta}{2\Gamma}\sinh(\Gamma z) \right)\NN\\
     V(z)  &=& e^{i \Delta z/2}  \frac{\sqrt{2} \alpha \lambda}{\Gamma}
     \sinh(\Gamma z) ,
     \EEA
where
  $\Gamma = \sqrt{2\alpha^2 \lambda^2-\frac{\Delta^2}{4}}$.
We shall be concerned for simplicity with the case of zero phase mismatch,
$\Delta=0$, and will be considering only the zeroth spatial Fourier
components of the field, $q=0$. Physically, this corresponds to
photodetection of the light field by a single large photodetector without
spatial resolution. In this case, we have $U(l)=\cosh r$ and $V(l)=\sinh r$
with $r=\sqrt{2} \alpha \lambda l$ being the squeezing parameter.

When the input states of the three
interacting modes are vacuum states,  the output
states remain Gaussian. Hence, they are completely described by the output covariance matrix $\sigma$
associated with the following quadrature components
 \BEA
 \hat{x}_0(\ell)&\equiv& 2 \re \hat{a}_0(\ell)\NN\\
\hat{p}_0(\ell) &\equiv&2 \im \hat{a}_0(\ell)\NN\\
\hat{x}_\pm(\ell)&\equiv&2 \re \hat{a}_\pm(\ell)\NN\\
\hat{p}_\pm(\ell) &\equiv&2 \im \hat{a}_\pm(\ell).
\EEA
Its matrix elements read
 \BE
 \sigma_{ij} = \textrm{Tr}\left[{\rho}\left(\Delta\hat{\xi}_i
 \Delta\hat{\xi}_j + \Delta\hat{\xi}_j \Delta\hat{\xi}_i\right)/2\right],
 \EE
where ${\rho}$ is the output density matrix and
$\Delta\hat{\xi}_i = \hat{\xi}_i-\left\langle\hat{\xi}_i\right\rangle_\rho$
with $\hat{\xi}_i$  the $i$-th component of the vector $\hat{{\bf
\xi}}=\left(\hat{x}_0(\ell),\hat{p}_0(\ell),\hat{x}_+(\ell),\hat{p}_+(\ell),\hat{x}_-(\ell),\hat{p}_-(\ell)\right)$.
Taking the solution (\ref{solution}) into account, the covariance matrix at
the output of the crystal reads explicitly
 \BE
   \sigma = \left(\begin{array}{cccccc}
   \cosh(2r) & 0 &\frac{ \sinh(2r) }{\sqrt{2}}& 0 &\frac{ \sinh(2r) }{\sqrt{2}}& 0 \\
   0 & \cosh(2r) & 0 & -\frac{ \sinh(2r) }{\sqrt{2}}& 0 & \frac{ -\sinh(2r) }{\sqrt{2}}  \\
  \frac{ \sinh(2r) }{\sqrt{2}}& 0 & \cosh^2r & 0 &\sinh^2r&0\\
 0& - \frac{ \sinh(2r) }{\sqrt{2}}& 0 & \cosh^2r & 0 &\sinh^2r\\
  \frac{ \sinh(2r) }{\sqrt{2}}& 0 & \sinh^2r & 0 &\cosh^2r&0\\
 0& - \frac{ \sinh(2r) }{\sqrt{2}}& 0 & \sinh^2r & 0 &\cosh^2r\\
   \end{array}\right) .      \label{sigma}
 \EE
 Notice that this matrix  is {\em bisymmetric} as it is invariant under the permutation of the quadratures  pertaining to the modes + and -: $\hat{x}_+(\ell),\hat{p}_+(\ell) \leftrightarrow \hat{x}_-(\ell),\hat{p}_-(\ell)$ . Hence, it has the tripartite entanglement structure of covariance matrices associated to bisymmetric (1+2)-mode Gaussian states considered in  Ref.~[3]. Actually, we can show that the output state obtained here exhibits genuine tripartite entanglement in the sense of Ref.~[20]. For that purpose, we have to verify that the following condition on covariance matrix elements is violated
 \BE
C \equiv \left\langle\left\{ \hat{x}_0(\ell)- \frac{\hat{x}_+(\ell)+\hat{x}_-(\ell)}{\sqrt 2}\right\}^2 \right\rangle_\rho +  \left\langle\left\{ \hat{p}_0(\ell)+ \frac{\hat{p}_-(\ell)+\hat{p}_-(\ell)}{\sqrt 2}\right\}^2 \right\rangle_\rho \geq \frac{1}{2}. \label{C}
 \EE
  From (\ref{sigma}) one deduces that 
  \BE
  C=4 \{\cosh (2r)-\sinh(2r)\}=4 e^{-2r},
  \EE 
  which is smaller than 1/2 if the squeezing parameter $r > \frac{3}{2} \ln 2 $. Recalling that $r=\sqrt{2} \alpha \lambda \ell$ therefore entails that the output state $\rho$ produced by the above parametric down-conversion process with two symmetrically-tilted
plane waves exhibits genuine tripartite entanglement when the pump amplitude $\alpha$, the coupling parameter $\lambda$ and the crystal length $\ell$ are such that
 \BE
 \alpha \lambda \ell > \frac{3 \ln 2}{2 \sqrt{2}} \approx 0.735.
 \EE
\subsection{Two-mode entanglement  localization}
We shall show that the tripartite entanglement generated by the above process can be concentrated on two modes by local unitary transformations as was discovered by Serafini, Adesso and Illuminati [3]. 
Let us consider the following quadratures
\BEA
\begin{array}{ll}
\hat{x}_0'(\ell)\equiv \hat{x}_0(\ell) & \hat{p}_0'(\ell)\equiv \hat{p}_0(\ell)\\
\hat{x}_1'(\ell)\equiv \frac{\hat{x}_+(\ell)+\hat{x}_-(\ell)}{\sqrt 2} &\hat{p}_1'(\ell)\equiv \frac{\hat{p}_+(\ell)+\hat{p}_-(\ell)}{\sqrt 2} \\
\hat{x}_2'(\ell)\equiv \frac{\hat{x}_+(\ell)-\hat{x}_-(\ell)}{\sqrt 2} & \hat{p}_2'(\ell)\equiv \frac{\hat{p}_+(\ell)-\hat{p}_-(\ell)}{\sqrt 2} ,
\end{array}
\EEA
which are  obtained from the previous ones by a unitary transformation corresponding to the action of a beamsplitter.
Accordingly, the output covariance matrix $\sigma$ is thus transformed by congruence,
\BE
\sigma' = S^{\rm T} \sigma S,
\EE
where $S$ is the following symplectic transformation,
\BEA
   S = I_1 \bigoplus \left(\begin{array}{cccc}
   \frac{1 }{\sqrt{2}}& 0 &\frac{1 }{\sqrt{2}}&0\\
 0 & \frac{1 }{\sqrt{2}}& 0 &\frac{1 }{\sqrt{2}}\\
  \frac{1 }{\sqrt{2}}& 0 &-\frac{1 }{\sqrt{2}}&0\\
 0 & \frac{1 }{\sqrt{2}}& 0 &-\frac{1 }{\sqrt{2}}
   \end{array}\right) ,     \label{S}
\EEA
 with $I_1$ the $2\times 2$ identity matrix. 
The new covariance matrix reads then
\BE
   \sigma' = \left(\begin{array}{cccc}
   \cosh(2r) & 0 & \sinh(2r) & 0 \\
   0 & \cosh(2r) & 0 & -\sinh(2r) \\
   \sinh(2r)& 0 & \cosh(2r) & 0 \\
   0 & -\sinh(2r) & 0 & \cosh(2r) \\
   \end{array}\right)\bigoplus I_1.
      \label{sigmaprim}
 \EE

The separability of the pertaining state
$\rho'$ can be determined by considering a partition of the system in two
subsystems and investigating the positivity of the partially transposed
matrix $\tilde \rho'$, obtained upon transposing the variables of only one
of the two subsystems. The positivity of partial transposition (PPT)  is a
necessary condition for the separability of any bipartite quantum
state [22], [23]. It is also sufficient for the separability
of $(1+n)-$mode Gaussian states [24], [25].

The covariance matrix $\tilde \sigma'$ of the partially transposed state
$\tilde \rho'$  with respect to one subsystem is obtained [24]  by
changing the signs of the quadratures  $\hat{p}_i$ belonging to that
subsystem.
Here, because of the block-diagonal structure of $\sigma'$ we
can restrict the analysis to the first block and consider the
transposition of the mode 1, i.~e.~change the sign of $\hat{p}_1'$,
\BE
  \tilde  \sigma' = \left(\begin{array}{cccc}
   \cosh(2r) & 0 & \sinh(2r) & 0 \\
   0 & \cosh(2r) & 0 & \sinh(2r) \\
   \sinh(2r)& 0 & \cosh(2r) & 0 \\
   0 & \sinh(2r) & 0 & \cosh(2r) \\
   \end{array}\right)\bigoplus I_1.
      \label{tildesigmaprim}
 \EE
The
symplectic eigenvalues $\{\tilde \nu_i\}$ of the covariance matrix $\tilde
\sigma'$ are $\tilde \nu_\pm =e^{\pm 2 r}$ and $1$. The necessary and
sufficient PPT condition for the separability of the state $\rho'$ amounts
to having $\tilde \nu_i \geq 1$ for all $i$. We can therefore focus on the
smallest eigenvalue $\tilde \nu_-$. The extent to which the separability
of the state is violated, is measured by $\mathcal{E_N}(\rho)$, the
logarithmic negativity of $\rho'$, defined as the logarithm of the trace
norm of $\tilde \rho'$,
 \BE
    \mathcal{E_N}(\rho')=\ln || \tilde \rho' ||_1=\max( 0, -\ln \tilde \nu_-).
 \EE
Using the calculated symplectic eigenvalues $\{\tilde \nu_i\}$ we obtain a
very simple expression for the logarithmic negativity,
 \BE
    \mathcal{E_N}(\rho')=2r.
 \EE
This implies that the central mode 0 is entangled with the mode 1 which is precisely the combination of modes + and - that was used in the condition (\ref{C}) in order to show the existence of genuine tripartite entanglement.

\section{Discussion and conclusion} 
Our scheme can be  generalized to multipartite entanglement localization using
$2N$ symmetrically tilted plane waves as pump modes in the parametric
down-conversion process. 
An illustration is provided in Fig. 2 for the case $N=2$ and 4 which correspond to 4 and 8 pump waves.   
\begin{center}
\begin{figure}
\includegraphics[scale=0.54]{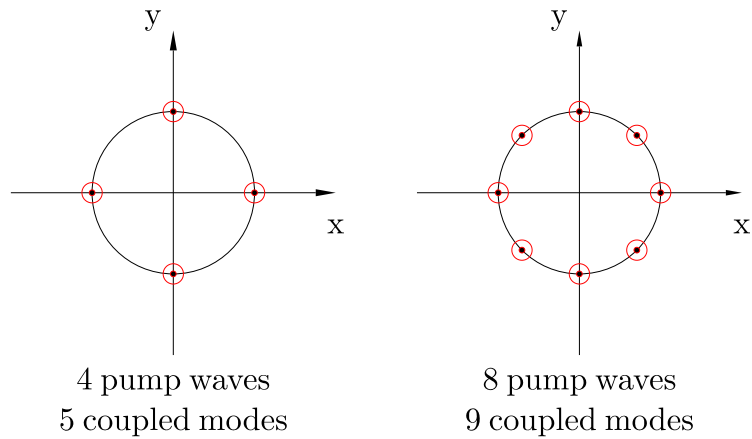}
\caption{(Color online)  Generalization of the scheme for $2N$ symmetrically tilted pump waves with $N=2$ and 4. The little circles represent the projections of the pump wave vectors in the $xy$ plane of the crystal entering face.}
\end{figure}
\end{center}
The projections of the pump wave vectors in the $xy$ plane of the crystal entering face are depicted as little circles.
In this case, the system contains $2N+1$ coupled modes which are described by equations similar to (\ref{3eq}).

 We can solve them in the rotating wave approximation
and  investigate the properties of the multipartite entanglement in
this case (the details will be provided in a forthcoming publication). The covariance matrix $\sigma$ at the output of the crystal is again bisymmetric as those considered in Ref.~[3]. By an orthogonal transformation of the field operators which generalizes the one achieved in the tripartite case,
 the covariance matrix
$\sigma$ can be brought to a form similar to (\ref{sigmaprim}), namely  the tensor product of 

 a $4 \times 4$ covariance matrix and a $2(2N-1)\times 2(2N-1)$ identity matrix. 
 The $4 \times 4$ covariance matrix has exactly the same form as the one in (\ref{sigma}) with $r$ replaced by $\sqrt{N} r$. 
Hence, it corresponds to that of two
entangled modes with the squeezing parameter enhanced by $\sqrt{N}$ and
$2N-1$ modes in the vacuum state. 
For the logarithmic
negativity we indeed obtain
$    \mathcal{E_N}(\rho')=2\sqrt{N}r$.
Physically, this means that by mixing
different spatial modes with beam splitters, we can ``localize'' the
entanglement distributed initially among all the $2N+1$ spatial modes
in only two well-defined modes, formed by the linear combinations of the
initial ones. Interestingly, this  entanglement localization results in an enhancement
of the entanglement by a factor $\sqrt{N}$.

To conclude, our ``active'' optical
scheme relies on the use of a spatially-structured pumping of the parametric medium. It
provides a realistic proposal for the experimental realization of multipartite entanglement as well as its localization.
The
possibility of entanglement localization was predicted mathematically in
Ref.~[3] and a  physical implementation was also given in terms of 2N-1 beam splitters and 2N single-mode squeezed inputs [20].
We have also established the scaling behavior of the logarithmic
negativity with the number of modes. Note that due to the asymmetric coupling between the modes in our
scheme, this scaling is a relatively slowly growing function of the
number of modes. An interesting question is to determine the ultimate rate
at which the logarithmic negativity may grow with the number of modes.
This subject will be addressed elsewhere.
 
 \section{Acknowledgments}
 
This work was supported by IRSIB (Institut d'encouragement de la Recherche
Scientifique et de l'Innovation de Bruxelles)  and the
FET programme under COMPAS FP7-ICT-212008 and HIDEAS FP7-ICT-221906.

\end{document}